# Model for quantum efficiency of guided mode plasmonic enhanced silicon Schottky detectors

Ilya Goykhman[1], Boris Desiatov[1], Joseph Shappir[1], Jacob B. Khurgin[2] and Uriel Levy[*,1]

[1]Department of Applied Physics, The Benin School of Engineering and Computer Science, The Center for Nanoscience and Nanotechnology, The Hebrew University of Jerusalem, Jerusalem, 91904, Israel.

[2]Department of Electrical and Computer Engineering, Johns Hopkins University, Baltimore Maryland 21218, USA

* Corresponding author: ulevy@cc.huji.ac.il

**Abstract**

Plasmonic enhanced Schottky detectors operating on the basis of the internal photoemission process are becoming an attractive choice for detecting photons with sub bandgap energy. Yet, the quantum efficiency of these detectors appears to be low compare to the more conventional detectors which are based on interband transitions in a semiconductor. Hereby we provide a theoretical model to predict the quantum efficiency of guided mode internal photoemission photodetector with focus on the platform of silicon plasmonics. The model is supported by numerical simulations and comparison to experimental results. Finally, we discuss approaches for further enhancement of the quantum efficiency.

**Introduction**

Silicon photodetectors are widely used in the visible spectral regime. Yet, such detectors cannot be used for the detection of radiation at the telecom band (e.g. 1.3-1.6 um wavelength) because the photon energy is lower than the energy bandgap of silicon (~1.1eV). One of the possibilities to overcome this limitation and to detect infrared sub-bandgap optical radiation in silicon is based on the internal photoemission (IPE). With this approach, one can realize a photo-detector by forming a Schottky contact at a metal-semiconductor interface. This Schottky type device is characterized by a potential barrier $\Phi_B$ between the Fermi level of the metal and the conduction band of the semiconductor, and shows rectifying electrical characteristics. If the energy of the photon is higher than this potential barrier, there is a probability for the hot electron generated in the metal (via excitation by a photon or by a plasmon) to overcome the barrier and arrive at the semiconductor. Unfortunately, the efficiency of this process is not high, as a result of the large momentum mismatch between the wavefunction of the electron in the metal in the semiconductor. Several models investigated the process in details, and provided predictions for the efficiency of the internal photoemission process [1-4]. These theories are based on the fundamental work of Fowler [5], which predicted that the efficiency is proportional to $\frac{(\hbar\omega - \Phi_B)^2}{\hbar\omega}$ where $\hbar\omega$ is the energy of the incident photon. A recent work was devoted to the prediction of the quantum efficiency of silicon plasmonic internal photoemission detectors [6]. This work assumed a finite escape cone for the electron in the metal. Within the cone, it was assumed that the transmission probability of the electron from the metal to the silicon is 100%, whereas outside of the cone the probability goes to 0. While the model provide a first order approximation to the behavior of the internal photoemission photodetectors, a more detailed model is required to account for the finite (and low!) transmission probability even within the escape cone. Furthermore, one should also take into account thermalization processes in the metal, which means that not all the electrons with are propagating towards the metal-semiconductor interface will make it before experiencing an inelastic collision. Hereby, we provide an extended model which takes into account these processes.

**Model**

The metal-semiconductor interface and the internal photoemission process is depicted in Fig. 1.

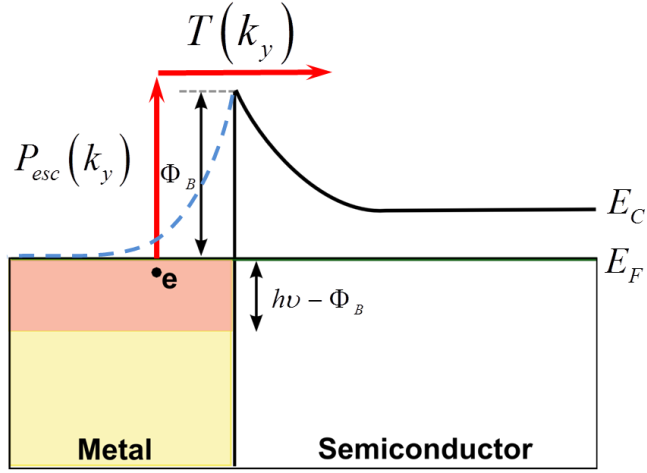

Fig. 1: Energy band diagram of the metal-semiconductor (n-type) Schottky interface and the internal photoemission process.

The photo detection process can be divided into two major steps: a) - transport of a hot electron from the metal to the Schottky interface, and b) – transmission of the electron through the interface from the metal to the semiconductor. We now discuss in details each component. For simplicity we assume a 1-d model.

a) – Transport of the hot electron in the metal to the Schottky interface

The probability of an electron to be excited by a plasmon is proportional to the intensity of the surface plasmon mode, which is decaying to the metal with a decay constant of $L_{spp}$. Thus, the probability of generating a hot electron by the photon (more accurately one should discuss the process of converting a plasmon to hot electron. Yet, in view of the frequencies involved, which are much below the plasma frequency, the plasmonic mode approaches the light line and thus the plasmon can be viewed as a photon), is given by:

$$P_g(x) = \frac{1}{L_{spp}} e^{-\frac{x}{L_{spp}}}$$

Once a hot electron has been generated, it has a finite probability of arriving at the interface. The electron may travel either towards the interface or away from it, thus one may expect it has a probability of 50% to arrive at the interface. Yet, the electron may experience thermalization during its transport. We assume that after a single inelastic collision the electron loose substantial energy and cannot contribute to the internal photoemission process.

The mean free path, i.e. the distance between collisions is given by $L_{MFP} = v_F \tau_D$ where $v_F$ is a Fermi velocity and $\tau_D$ is the relaxation time. The probability of an electron which has been excited at a distance x away from the interface to reach the interface is given by:

$$P_{MFP} = e^{-\frac{x}{L_{MFP}}}$$

Here we assume that the electron is propagating towards the interface at normal incident angle. While this is only an approximation, we later show that it is a reasonable assumption. Combining the previous two expressions, once can calculate the electron transport probability F, i.e. the probability of a hot electron excited by SPP wave to reach the interface before thermalization occurs:

$$F = \int_0^\infty P_{spp}\, P_{MFP}\, dx = \frac{1}{L_{spp}} \int_0^\infty e^{-\frac{x}{L_{spp}}} e^{-\frac{x}{L_{MFP}}} dx = \frac{L_{MFP}}{L_{MFP} + L_{spp}} = \frac{1}{1 + L_{spp}/L_{MFP}}$$

b) Electron transmission through the metal-silicon interface

After being transported to the interface, the hot electron has a finite probability of crossing towards the semiconductor. This is because of the huge difference in its momentum in the metal and in the semiconductor. We next study this probability in more details.

First we consider the energy-momentum relation of an electron in the semiconductor, based on the parabolic approximation of effective mass. The energy of the transmitted hot electron relative to the bottom of conduction band ($E_C$) is related to its momentum by:

$$E - \Phi_B = \frac{\hbar^2 K_S^2}{2m_e^*} = \frac{\hbar^2}{2m_e^*}\left(K_{S,x}^2 + K_{S,y}^2\right) \quad for\ \Phi_B < E < \hbar\omega$$

$$K_S^2 = K_{S,x}^2 + K_{S,y}^2$$

where $K_{s,x}$ and $K_{s,y}$ are the normal and the parallel components of the wave vector in the semiconductor with respect to the interface and $m_e^*$ is the effective mass of an electron in the semiconductor. For simplicity, we neglect the difference in effective mass for the parallel and the transverse direction.

In the metal, the energy of the excited hot electron is given by:

$$E_M = E_F + E = \frac{\hbar^2 K_m^2}{2m_e} = \frac{\hbar^2}{2m_e}\left(K_{m,x}^2 + K_{m,y}^2\right)$$

$$K_m^2 = K_{m,x}^2 + K_{m,y}^2$$

where $K_{m,x}$ and $K_{m,y}$ are the normal and the parallel components of the wave vector in the metal with respect to the interface. We now force the continuity of the parallel wave vector component at the interface, i.e.

$$K_{s,y} = K_{m,y} = K_y$$

This requirement imposes an escape cone (in the K-space) for electrons in the metal. This is because For each particular electron energy E in the range of interest $\Phi_B < E < \hbar\omega$, the parallel component of the wave vector is distributed over the interval $0 < K_y(E) < K_s(E)$, where $\max[K_y(E)] = \max[K_s(E)] = \sqrt{\left[\frac{2m_e^*(E-\Phi_B)}{\hbar^2}\right]}$ defines the escape cone with a solid angle $\Theta_{max}(E)$. Using realistic Schottky barrier values for silicon and standard plasmonic metals (e.g. Au, Ag or Al), and assuming the wavelength of interest around 1.5 microns, the angular escape cone is very small. Therefore, considering only normal velocity direction in the probability for thermalization seems to be a reasonable approximation. The escape cone is depicted schematically in Fig. 2.

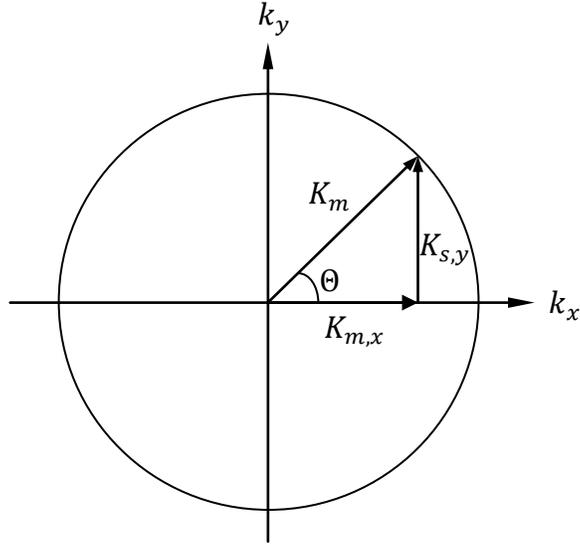

Fig. 2: Schematic description of the escape cone.

Outside the escape cone, the probability for an electron to be transmitted from the metal to the semiconductor approaches zero. Yet, even within the escape cone the transmission probability is relatively low. We now express the transmission probability for a given energy E , as a function of the parallel wave vector component in the interval $0 < K_y(E) < K_s(E)$ . The expression is similar to the Fresnel transmission coefficient of light crossing between two media, an expression which is widely used in photonics:

$$T(K_y) = 1 - \left(\frac{K_{m,x} - K_{s,x}}{K_{m,x} + K_{s,x}}\right)^2 = \frac{4 K_{m,x} K_{s,x}}{(K_{m,x} + K_{s,x})^2} = \frac{4 * \sqrt{(K_m^2 - K_y^2)(K_s^2 - K_y^2)}}{\left(\sqrt{K_m^2 - K_y^2} + \sqrt{K_s^2 - K_y^2}\right)^2}$$

For a given value of $K_y(E)$ we can associate an angle of propagation according to:

$$sin\theta = \frac{K_y}{K_m} \;\; ; \;\; cos\theta = \frac{K_{m,x}}{K_m} = \frac{\sqrt{K_m^2 - K_y^2}}{K_m} \;\; ; \;\; dK_y = K_m cos\theta d\theta$$

The solid angle related to this angle is given by: $d\Omega(\theta) = sin\theta d\theta$. We can express $d\Omega$ as a function of $K_y$:

$$d\Omega(K_y) = \frac{K_y}{K_m} * \frac{dK_y}{K_m cos\theta} = \frac{K_y}{K_m} * \frac{dK_y}{K_{m,x}} = \frac{K_y dK_y}{K_m \sqrt{K_m^2 - K_y^2}}$$

For a given value of $K_y$, the probability of an electron at the interface to be transmitted is proportional $T(K_y)d\Omega(K_y)$. Thus, the overall transmission probability, for a given energy E is given by:

$$P(E) = \frac{1}{4\pi} \int_0^{2\pi} d\emptyset \int_0^{K_s(E)} d\Omega(K_y) T(K_y) dK_y$$

Substituting $d\Omega(K_y)$ and $T(K_y)$, one obtains:

$$P(E) = \int_0^{K_s(E)} \frac{2\sqrt{(K_s^2 - K_y^2)}}{\left(\sqrt{K_m^2 - K_y^2} + \sqrt{K_s^2 - K_y^2}\right)^2} \frac{K_y dK_y}{K_m}$$

The energy of the hot electrons is distributed over a range of energies. This is because the electron may be excited not only from the Fermi surface, but also from within the Fermi sphere. Thus, one needs to find the overall transmission probability by performing an integration on the probability for each energy value that is sufficient for supporting the photoemission process. This overall probability is given by:

$$P = \frac{1}{\hbar\omega - \Phi_B} \int_{\Phi_B}^{\hbar\omega} P(E) \, dE$$

We can neglect the dependency of the density of states in energy because the range of energies $\hbar\omega - \Phi_B$ is small as compared to the Fermi energy. Finally, the overall probability, for an excited electrons in energy interval $\Phi_B < E < \hbar\omega$ to reach the interface without thermalization and to be injected into the semiconductor is given by:

$$P_{tot} = P * F$$

This expression is the internal quantum efficiency (IQE) of the device.

**Numerical simulations**

After providing this elaborated model for the prediction of the internal photoemission probability, we turn into evaluating a simple case study, where a surface plasmon mode is propagating along a silicon-metal interface. We solve for the surface plasmon propagation constant in order to find $L_{spp}$, and study the quantum efficiency of the internal photoemission process as a function of the Schottky Barrier, wavelength of light and Fermi energy. The results are shown in Fig. 3,4. As expected, the IQE increases with the decrease in wavelength and the decrease in Schottky barrier (because of the higher photon energy with respect to the barrier). Also, we note that the IQE is increasing with the decrease of the Fermi energy. This is because the momentum mismatch is reduced for lower values of Fermi energy. These trends has been qualitatively observed in [7,8]. Yet, the measured IQE values were in general higher than those expected by our model. Possible explanations may be related to additional injection mechanisms such as tunneling, as well as to roughness at the interface which results in a relaxation of the momentum matching condition. Furthermore, while our model is purely one dimensional, the actual devices reported in [7, 8] are two dimensional and thus provide additional routes for electron injection.

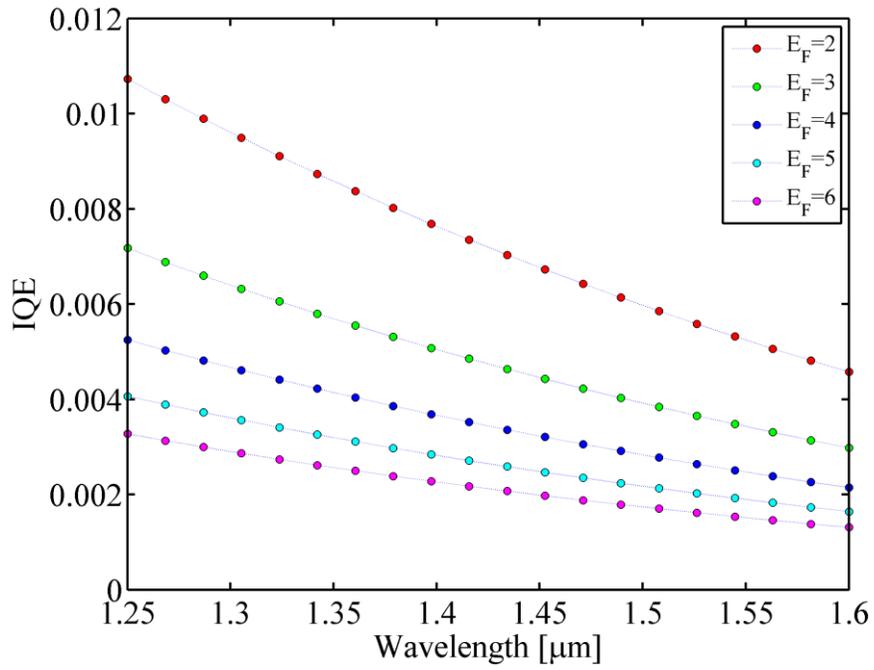

Fig. 3: IQE as a function of the exciting wavelength for different values of the Fermi energy and the Schottky barrier=0.55eV.

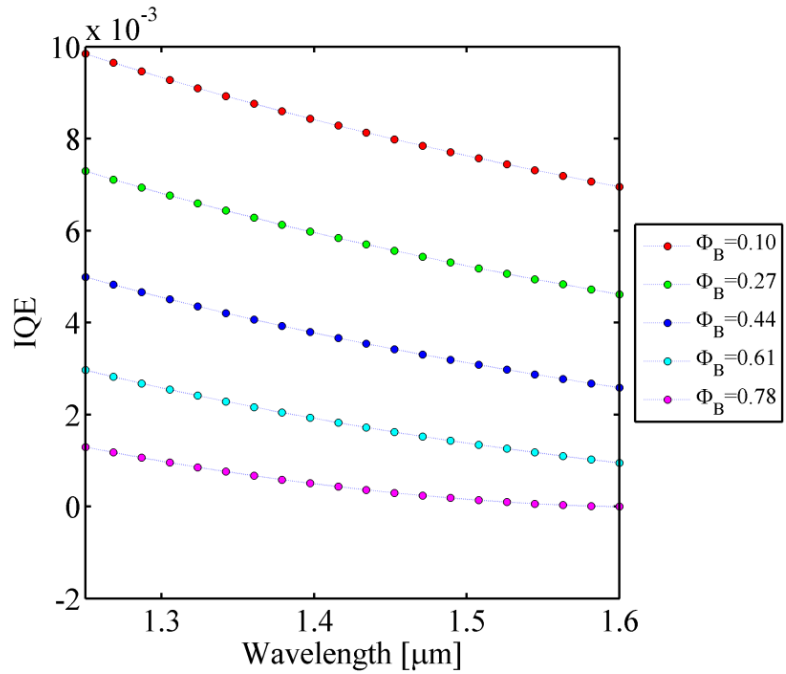

Fig. 4: IQE as a function of the exciting wavelength for different values of the Schottky barrier and the Fermi energy=5.5eV.

**References**

1. J. E. Sipe and J. Becher, "Surface-plasmon-assisted photoemission," J. Opt. Soc. Am. 71(10), 1286-1288 (1981).
2. Mooney, J. M. and J. Silverman ,"The theory of hot-electron photoemission in schottky-barrier IR detectors." Electron Devices, IEEE Transactions on 32 (1), 33-39 (1985).
3. Scales, C. and P. "Berini Thin-Film schottky barrier photodetector models," Quantum Electronics, IEEE Journal of 46 (5), 633-643 (2010).
4. Ivanov, V. G. "Quantum efficiency of schottky photodiodes near the long-wavelength edge," Semiconductors 31 (6), 631+ (1997)
5. R. H. Fowler, "The analysis of photoelectric sensitivity curves for clean metals at various temperatures," Phys. Rev. 38 (1), 45–56 (1931).
6. Scales, C., I. Breukelaar, R. Charbonneau, and P. Berini (2011, June). Infrared performance of symmetric Surface-Plasmon waveguide schottky detectors in si. Lightwave Technology, Journal of 29 (12), 1852-1860.
7. I. Goykhman, B. Desiatov, J. Khurgin, J. Shappir, and U. Levy," `Locally Oxidized Silicon Surface-Plasmon Schottky Detector for Telecom Regime'. *Nano Lett.* 11(6)", Opt. Express 20, 28594 (2011)
8. I. Goykhman, B. Desiatov, J. Khurgin, J. Shappir, and U. Levy, " Waveguide based compact silicon Schottky photodetector with enhanced responsivity in the telecom spectral band" , Opt. Express 20, 28594 (2012)